\documentstyle[12pt]{article}

\newcommand {\vs}[1]  { \vspace*{#1 cm} }

\newcounter{eq}
\newcounter{sc}

\newcommand {\MPL}  {Mod. Phys. Lett.}
\newcommand {\IJMP}  {Int. J. Mod. Phys.}
\newcommand {\NP}   {Nucl. Phys.}

\newcommand {\PL}   {Phys. Lett.}

\newcommand {\PR}   {Phys. Rev.}

\newcommand {\AP}   {Ann. of Phys.}

\def\overleftrightarrow#1{\vbox{\ialign{##\crcr
 $\leftrightarrow$\crcr\noalign{\kern-1pt\nointerlineskip}
 $\hfil\displaystyle{#1}\hfil$\crcr}}}

\newlength{\minitwocolumn}
\begin{document}

\begin{flushright}
DPUR/TH/20\\
March, 2010\\
\end{flushright}
\vspace{30pt}
\pagestyle{empty}
\baselineskip15pt

\begin{center}
{\large\bf Higgs Mechanism for Gravitons

 \vskip 1mm
}

\vspace{10mm}

Ichiro Oda
          \footnote{
           E-mail address:\ ioda@phys.u-ryukyu.ac.jp
                  }

\vspace{10mm}
          Department of Physics, Faculty of Science, University of the 
           Ryukyus,\\
           Nishihara, Okinawa 903-0213, JAPAN \\

\end{center}


\vspace{10mm}
\begin{abstract}
Just like the vector gauge bosons in the gauge theories, it is now
known that gravitons acquire mass in the process of spontaneous symmetry
breaking of diffeomorphisms through the condensation of scalar fields.
The point is that we should find the gravitational Higgs mechanism such that
it results in massive gravity in a flat Minkowski space-time without non-unitary
propagating modes. This is usually achieved by including higher-derivative terms 
in scalars and tuning the cosmological constant to be a negative value in a 
proper way. Recently, a similar but different gravitational Higgs mechanism
has been advocated by Chamseddine and Mukhanov where one can relax the negative
cosmological constant to zero or positive one.
In this work, we investigate why the non-unitary ghost mode decouples from
physical Hilbert space in a general space-time dimension. Moreover, we
generalize the model to possess an arbitrary potential and clarify under what conditions
the general model exhibits the gravitational Higgs mechanism.
By searching for solutions to the conditions, we arrive at two classes of potentials 
exhibiting gravitational Higgs mechanism. One class includes the model by Chamseddine 
and Mukhanov in a specific case while the other is completely a new model.

\vspace{15mm}

\end{abstract}

\newpage
\pagestyle{plain}
\pagenumbering{arabic}


\rm
\section{Introduction}

Recently, interests on the construction of massive gravity theories have
revived from different physical motivations \cite{Percacci}-\cite{Rubakov2}. 
A fundamental question in these studies is how one can make theories of
massive gravity with a small but finite graviton mass.  By assuming the mass
to be very tiny, the gravitational interaction deviates from the predictions
of Einstein's general relativity only at large scales which are comparable to
the Compton wave length of gravitons.

One motivation behind these interests comes from the astonishing 
observational fact that our universe is not just expanding but is at present 
in an epoch of undergoing an accelerating expansion \cite{Riess, Perlmutter, WMAP}. 
Although the standard model of cosmology based on general relativity is not only 
remarkably successful in accounting for many of observational facts of the 
universe but also the study of general relativity has matured into precise science 
with many impressive observable tests which include a familiar example like 
its use of the global positioning system (GPS), we do not yet have a firm grasp of 
the late-time cosmic acceleration in addition to problems associated with dark 
matter and dark energy. 

Massive gravity theories might play a role in the sense that they could modify 
Einstein's general relativity at large cosmological scales and might lead to 
the present accelerated expansion of the universe without assuming still mysterious 
dark matter and dark energy. It is worthwhile to point out the fact that 
since general relativity is almost the unique theory of massless spin 2 gravitational 
field whose universality class is determined by local symmetries under diffeomorphisms, 
any infrared modification of general relativity cannot help introducing some kind of 
mass for gravitons.

The other motivation for attempting to construct massive gravity theories
is conceptual and is related to the noncritical string theory applied to 
quantum chromodynamics (QCD) \cite{'t Hooft}.
For instance, as inspired in the large-N expansion of the gauge theory,
which defines the planar (or genus zero) diagram and is analogous to the tree
diagram of string theory, if we wish to apply a bosonic string theory 
to the gluonic sector in QCD, massless fields such as spin 2 graviton
in string theory, must become massive or be removed somehow by an ingenious 
dynamical mechanism since such the massless fields do not appear in QCD.
Note that this motivation is relevant to the modification of general relativity
at small scales whereas the previous cosmological one is to the 
infrared modification. 

A few years ago, 't Hooft proposed a new Higgs mechanism for gravitons where 
the massless gravitons '$\it{eat}$' four real scalar fields and consequently
become massive \cite{'t Hooft}. In his model, vacuum expectation values (VEV's) 
of the scalar fields are taken to be the four space-time coordinates by gauge-fixing 
diffeomorphisms, so the whole diffeomorphisms are broken spontaneously. 
Of course, the number of dynamical degrees of freedom should be left unchanged 
before and after the SSB. Actually, before the SSB of diffeomorphisms 
there are massless gravitons of two dynamical degrees of freedom and 
four real scalar fields whereas after the SSB we have massive gravitons of five 
dynamical degrees of freedom and one real scalar field. 
Afterward, a topological term was included to the 't Hooft model 
where an '$\it{alternative}$' metric tensor is naturally derived and the topological 
meaning of the gauge conditions was clarified \cite{Oda}. \footnote{Similar but 
different approaches have been already taken into consideration in Ref. \cite{Oda1}.}  

One serious problem in the 't Hooft model is that a scalar field appearing 
after the SSB is a non-unitary propagating field so that in order to keep the
unitarity the non-unitary mode must be removed from the physical Hibert space 
in terms of some procedure. This problem was solved by including higher-derivative 
terms in the scalar fields and tuning appropriately the cosmological constant 
to be a negative value in Ref. \cite{Kaku2}. 

More recently, Chamseddine and Mukhanov have presented a new Higgs mechanism for
gravitons also by adding higher-derivative terms in scalars to the Einstein-Hilbert
action \cite{Chamseddine}. One advantage of their model is that we do not have to
restrict the cosmological constant to be negative, namely zero or positive cosmological
constant is also allowed to trigger the gravitational Higgs mechanism.

The aims of this article are three-fold. First, we simply generalize the model
by Chamseddine and Mukhanov to a general $D$-dimensional space-time. Second,
we will present a proof that the model does not involve the ghost so it is unitary.
This proof is performed by using the gauge-variant fluctuations and fixing diffeomorphisms
in two steps. The key point for decoupling the ghost is the existence of
residual gauge symmetry like that of QED in the Lorentz gauge. Finally, we construct
a general model with an arbitrary function of $H^{AB}$ and find the conditions
by requiring that the model should exhibit the Higgs mechanism for gravitons
in a flat Minkowski background. It is remarkable that there exist two kinds of 
solutions to the conditions. One of them includes the model by Chamseddine and Mukhanov
while the other is completely a new class of models. We also present a specific
model of the latter.

This paper is organized as follows: In section 2, we review the model proposed by 
Chamseddine and Mukhanov.  In section 3, we present a proof of unitarity of the model
by fixing diffeomorphisms by gauge conditions in two steps. In section 4, we construct
a more general model and search for new models showing the Higgs mechanism of gravitons.
The final section is devoted to conclusions and discussion.

\section{Review on the Chamseddine and Mukhanov model}

In this section, we wish to review on the model by Chamseddine and Mukhanov 
\cite{Chamseddine} in a general $D$-dimensional space-time.
Consider the following action \footnote{We obey the conventions and the notation 
in the Misner et al.'s textbook \cite{MTW}.}:
\begin{eqnarray}
S = \frac{1}{16 \pi G} \int d^D x \sqrt{-g} [ R + \frac{m^2}{4} 
\{ \frac{D(D-1)}{4} ( (\frac{1}{D} H)^2 - 1 )^2 
- \tilde H_{AB} \tilde H^{AB} \} ].
\label{CM model}
\end{eqnarray}
Here $G$ is the $D$-dimensional Newton's constant and the induced internal metric 
$H^{AB}$ is defined as
\begin{eqnarray}
H^{AB} = g^{\mu\nu} \nabla_\mu \phi^A \nabla_\nu \phi^B,
\label{H}
\end{eqnarray}
where $\phi^A$ are real $D$ scalar fields with $A = 0, \cdots, D-1$.
And $H$ and $\tilde H^{AB}$ are respectively the trace and the traceless part
of $H^{AB}$ defined by
\begin{eqnarray}
H^{AB} = \tilde H^{AB} + \frac{1}{D} \eta^{AB} H,
\label{tilde H}
\end{eqnarray}
where the indices $A, B, \cdots$ are raised and lowered in terms of
the Minkowski metric $\eta_{AB} = diag(-1, 1, \cdots, 1)$.

The equations of motion read
\begin{eqnarray}
&{}& \nabla^\mu [ \{ \tilde H^{AB} - \frac{D-1}{2D} \eta^{AB}
( (\frac{1}{D} H)^2 - 1 ) H \} \nabla_\mu \phi_B ]  = 0,
\nonumber\\
&{}& R_{\mu\nu} - \frac{1}{2} g_{\mu\nu} R 
= \frac{m^2}{8} g_{\mu\nu} [ \frac{D(D-1)}{4} \{ (\frac{1}{D} H)^2 - 1 \}^2 
- \tilde H_{AB} \tilde H^{AB} ]
\nonumber\\
&+& \frac{m^2}{4} [ - \frac{D-1}{D} \{ (\frac{1}{D} H)^2 - 1 \} 
H \nabla_\mu \phi^A \nabla_\nu \phi_A
+ 2 \tilde H^{AB} \nabla_\mu \phi_A \nabla_\nu \phi_B ].
\label{Eq.1}
\end{eqnarray}

We are interested in obtaining 'vacuum' solution of the form
\begin{eqnarray}
\phi^A &=& x^\mu \delta_\mu^A, 
\nonumber\\
g_{\mu\nu} &=& \eta_{\mu\nu}.
\label{Background}
\end{eqnarray}
This vacuum solution is not static since one component of $\phi^A$, that is,
$\phi^0$ is essentially equivalent to time $x^0 = t$.
Indeed, it is easy to check that the equations of motion (\ref{Eq.1}) have
the solution (\ref{Background}) as a classical solution.

We now expand the fields around this vacuum as 
\begin{eqnarray}
\phi^A &=& x^\mu \delta_\mu^A + \varphi^A, 
\nonumber\\
g_{\mu\nu} &=& \eta_{\mu\nu} + h_{\mu\nu},
\label{Fluctuation}
\end{eqnarray}
and write out all the terms up to second order, for instance
\begin{eqnarray}
H^{AB} = \eta^{AB} - \bar{h}^{AB} + \cdots,
\label{H-fluctuation}
\end{eqnarray}
where the ellipses stand for quadratic and higher order terms, note the minus sign 
in front of the second term owing to 
$g^{\mu\nu} = \eta^{\mu\nu} - h^{\mu\nu} + h^{\mu\alpha} h_{\alpha}^\nu 
+ \cdots$, and we have defined $\bar{h}^{AB} \equiv h^{AB} 
- \partial^A \varphi^B - \partial^B \varphi^A$.
All the indices are now raised and lowered by $\eta_{AB}$ and $\eta_{\mu\nu}$,
so the difference between the space-time $\mu$ and the internal $A$ indices
is not essential.

Next, let us consider diffeomorphisms in the infinitesimal forms:
\begin{eqnarray}
\delta \varphi^A &=& \xi^\mu \nabla_\mu \phi^A \approx \xi^A,
\nonumber\\
\delta h_{\mu\nu} &=& \nabla_\mu \xi_\nu +  \nabla_\nu \xi_\mu
\approx \partial_\mu \xi_\nu +  \partial_\nu \xi_\mu.
\label{Diffeo}
\end{eqnarray}
Under diffeomorphisms, it turns out that the $\bar{h}^{AB}$ is invariant.

The linearized equations of motion have the form:
\begin{eqnarray}
&{}& \partial^\nu \bar{h}_{\mu\nu} - \partial_\mu \bar{h} = 0,
\nonumber\\
&{}& \Box \bar{h}_{\mu\nu} + \partial_\mu \partial_\nu \bar{h} 
- \partial_\mu \partial_\rho \bar{h}_\nu^\rho
- \partial_\nu \partial_\rho \bar{h}_\mu^\rho - \eta_{\mu\nu} ( \Box \bar{h} 
- \partial_\rho \partial_\sigma \bar{h}^{\rho\sigma}) = m^2 ( \bar{h}_{\mu\nu}
- \eta_{\mu\nu} \bar{h} ),
\label{Linearized Eq.1}
\end{eqnarray}
where $\Box \equiv \eta^{\mu\nu} \partial_\mu \partial_\nu$.
Taking the trace of the latter equation, using the former equation, and
assuming $D \ne 1$, we obtain
\begin{eqnarray}
\bar{h} = 0.
\label{Linearized Eq.2}
\end{eqnarray}
With this equation, the former equation reduces to the form
\begin{eqnarray}
\partial^\nu \bar{h}_{\mu\nu} = 0.
\label{Linearized Eq.3}
\end{eqnarray}
Together with (\ref{Linearized Eq.2}) and (\ref{Linearized Eq.3}), Einstein's equations
in (\ref{Linearized Eq.1}) read 
\begin{eqnarray}
( \Box - m^2 ) \bar{h}_{\mu\nu} = 0.
\label{Linearized Eq.4}
\end{eqnarray}
Then, Eq's. (\ref{Linearized Eq.2})-(\ref{Linearized Eq.4}) have the same form
as those of massive gravity of Fierz-Pauli type \cite{Fierz} except that now
the equations are entirely written by not gauge-variant $h_{\mu\nu}$ but gauge-invariant 
$\bar{h}_{\mu\nu}$ including the scalar fluctuations $\varphi^A$. 
Note that the trace $\bar{h}$, which is nothing but the ghost mode, decouples
because of Eq. (\ref{Linearized Eq.2}).

\section{Proof of the unitarity}

Although it was found that the ghost mode decouples from physical modes as shown in the 
previous section, there remain some questions. First, it is not customary that the linearized 
equations of motion (\ref{Linearized Eq.2})-(\ref{Linearized Eq.4}) are written in terms of 
the gauge-invariant objects $\bar{h}_{\mu\nu}$ in which we have the scalar fluctuations 
$\varphi^A$ of $D$ degrees of freedom.  Second, at first sight, it seems that the number
of dynamical degrees of freedom is in conflict with the following naive counting of the 
degrees of freedom. Before the SSB of diffeomorphisms, we have massless gravitons of $\frac{D(D-3)}{2}$ 
and real scalars of $D$ components, one of which, $\varphi^0$, is timelike. On the other hand,
after the SSB, we would have massive gravitons of $\frac{(D+1)(D-2)}{2}$ degrees of freedom.
The difference of degrees of freedom between before and after the SSB is therefore
\begin{eqnarray}
\frac{D(D-3)}{2} + D - \frac{(D+1)(D-2)}{2} = 1,
\label{Counting}
\end{eqnarray}
which corresponds to the number of ghost. Thus, a simple but important question is
why the ghost $\it{does}$ decouple. In order to resolve this question, it was shown
in Ref. \cite{Chamseddine} that assuming vanishing gravitons $h_{\mu\nu} = 0$,
the kinetic term for the scalar fluctuations is reorganized into that of the gauge 
fields where we have an extra gauge symmetry for killing the ghost.  

In this paper, we shall present the different line of arguments for proving 
the unitarity, in other words, decoupling of the ghost mode, without assuming 
the vanishing gravitational fields. For this purpose, the key idea is to 
rewrite the whole equations of motion (\ref{Linearized Eq.1}) 
by the gauge-variant fields $h_{\mu\nu}$ and 
$\varphi^A$ instead of the gauge-invariant ones $\bar{h}_{\mu\nu}$. 
By making use of the definition
$\bar{h}^{AB} \equiv h^{AB} - \partial^A \varphi^B - \partial^B \varphi^A$, 
the result reads
\begin{eqnarray}
&{}& \partial^\nu h_{\mu\nu} - \partial_\mu h - \Box \varphi_\mu
+ \partial_\mu \partial_\nu \varphi^\nu = 0,
\nonumber\\
&{}& \Box h_{\mu\nu} + \partial_\mu \partial_\nu h 
- \partial_\mu \partial_\rho h_\nu^\rho
- \partial_\nu \partial_\rho h_\mu^\rho - \eta_{\mu\nu} ( \Box h 
- \partial_\rho \partial_\sigma h^{\rho\sigma}) 
\nonumber\\
&=& m^2 [ h_{\mu\nu} - \partial_\mu \varphi_\nu - \partial_\nu \varphi_\mu
- \eta_{\mu\nu} ( h - 2 \partial_\rho \varphi^\rho ) ].
\label{Linearized Eq.1-2}
\end{eqnarray}

Next, we attempt to fix diffeomorphisms by gauge conditions in two steps \footnote{This 
technique was also employed in Ref. \cite{Kaku2}.}. We first take the following gauge conditions
for only $D-1$ diffeomorphisms:
\begin{eqnarray}
\varphi_\mu = \partial_\mu \omega.
\label{Gauge1}
\end{eqnarray}
Then, the former equation in Eq. (\ref{Linearized Eq.1-2}) reduces to the form
\begin{eqnarray}
\partial^\nu h_{\mu\nu} - \partial_\mu h = 0.
\label{Linearized Eq.1-2-1}
\end{eqnarray}
With the help of Eq's. (\ref{Gauge1}) and (\ref{Linearized Eq.1-2-1}), the latter
equation in (\ref{Linearized Eq.1-2}) reads
\begin{eqnarray}
\Box h_{\mu\nu} - \partial_\mu \partial_\nu h 
= m^2 [ h_{\mu\nu} - 2 \partial_\mu \partial_\nu \omega 
- \eta_{\mu\nu} ( h - 2 \Box \omega ) ].
\label{Linearized Eq.1-2-2}
\end{eqnarray}
Taking the trace of this equation, we obtain
\begin{eqnarray}
h - 2 \Box \omega = 0,
\label{Linearized Eq.1-2-3}
\end{eqnarray}
by which Eq. (\ref{Linearized Eq.1-2-2}) is further simplified to be
\begin{eqnarray}
(\Box - m^2) h_{\mu\nu} = \partial_\mu \partial_\nu \hat{h},
\label{Linearized Eq.1-2-4}
\end{eqnarray}
where we have defined $\hat{h} \equiv h - 2 m^2 \omega$.

Next, let us take into consideration one diffeomorphism remaining, which is written as
\begin{eqnarray}
\xi_\mu = \partial_\mu \lambda.
\label{Gauge parameter}
\end{eqnarray}
Under this remaining diffeomorphism, each field is transformed like
\begin{eqnarray}
\delta \omega &=& \lambda,
\nonumber\\
\delta h_{\mu\nu} &=& 2 \partial_\mu \partial_\nu \lambda,
\nonumber\\
\delta h &=& 2 \Box \lambda,
\nonumber\\
\delta \hat{h} &=& 2 ( \Box - m^2 ) \lambda
\label{Diffeo remaining}
\end{eqnarray}
Using this remaining diffeomorphism, we can take the gauge condition
\begin{eqnarray}
\hat{h} = 0.
\label{Gauge2}
\end{eqnarray}
With this gauge condition, Eq. (\ref{Linearized Eq.1-2-4}) becomes the massive
Klein-Gordon equation for $h_{\mu\nu}$.

However, at this stage, it turns out that there remains a residual diffeomorphism. Actually,
under the remaining diffeomorphism satisfying the equation
\begin{eqnarray}
( \Box - m^2 ) \lambda = 0,
\label{Residual Diffeo}
\end{eqnarray}
$\hat{h}$ is obviously invariant as seen in Eq. (\ref{Diffeo remaining}). Note that
this equation is the same as that for gravitons \footnote{This situation is very
similar to that of QED in the Lorentz gauge where even after selecting the
Lorentz gauge condition $\partial_\mu A^\mu = 0$, there exists a residual
gauge symmetry $\delta A_\mu = \partial_\mu \Lambda$ with $\Lambda$ satisfying 
$\Box \Lambda = 0$, thereby further removing the longitudinal component of $A_\mu$.}.  
The existence of this residual gauge symmetry makes it possible to take the gauge 
\begin{eqnarray}
h = 0,
\label{Gauge3}
\end{eqnarray}
since $\delta h = 2 \Box \lambda = 2 m^2 \lambda$.

Consequently, the equations of motion for the gravitational fields take the form
\begin{eqnarray}
( \Box - m^2 ) h_{\mu\nu} &=& 0,
\nonumber\\
\partial^\nu h_{\mu\nu} &=& 0,
\nonumber\\
h &=& 0.
\label{Eq for gravitons}
\end{eqnarray}
This precisely describes massive gravitons of $\frac{(D+1)(D-2)}{2}$ degrees of freedom
without the ghost mode \cite{Fierz}.

\section{A more general model}

In this section, we discuss a generalization of the Chamseddine and Mukhanov model.
It is natural to generalize the action (\ref{CM model}) as follows:
\begin{eqnarray}
S = \frac{1}{16 \pi G} \int d^D x \sqrt{-g} [ R - V(H^{AB}) ],
\label{Generalized model}
\end{eqnarray}
where $\it{a \ priori}$ $V$ is a generic function of $H^{AB}$.

This general action gives us the following equations of motion:
\begin{eqnarray}
\nabla^\mu ( \frac{\partial V}{\partial H^{AB}} \nabla_\mu \phi^B ) &=& 0,
\nonumber\\
R_{\mu\nu} - \frac{1}{2} g_{\mu\nu} R 
&=& - \frac{1}{2} g_{\mu\nu} V + \nabla_\mu \phi^A \nabla_\nu \phi^B 
\frac{\partial V}{\partial H^{AB}}.
\label{Generalized Eq.1}
\end{eqnarray}

As in the Chamseddine and Mukhanov model, we are interested in the situation where
the vacuum (\ref{Background}) is allowed to be a classical solution to 
(\ref{Generalized Eq.1}). The requirement of the presence of the vacuum solution 
leads to a constraint on the potential $V$ such that the equation
\begin{eqnarray}
\frac{\partial V(H_*)}{\partial H^{AB}} = \frac{1}{2} \eta_{AB} V(H_*),
\label{Potential}
\end{eqnarray}
should be fulfilled where we have defined $H^{AB}_* = \eta^{AB}$. In other words,
the equation (\ref{Potential}) is a constraint imposed on the potential $V$ in order to
have a flat Minkowski space-time as the background. 

For instance, in the Chamseddine and Mukhanov model, the potential is given by
\begin{eqnarray}
V = - \frac{m^2}{4} [ \frac{D(D-1)}{4} \{ (\frac{1}{D} H)^2 - 1 \}^2 
- H_{AB} H^{AB} + \frac{1}{D^2} H^2 ],
\label{CM Potential}
\end{eqnarray}
so the equation (\ref{Potential}) is trivially satisfied since
\begin{eqnarray}
\frac{\partial V(H_*)}{\partial H^{AB}} = \frac{1}{2} \eta_{AB} V(H_*) = 0.
\label{CM Potential2}
\end{eqnarray}

As before, we expand the fields around the vacuum solution (\ref{Background})
as in (\ref{Fluctuation}).  Now we gauge away the scalar fluctuations
$\varphi^A$ by using diffeomorphisms. Of course, once we gauge away the
$D$ scalars, we can no longer gauge away any components of the gravitational
fluctuations $h_{\mu\nu}$. 

After setting $\varphi^A = 0$, the linearized equations of motion for (\ref{Generalized Eq.1})
read
\begin{eqnarray}
&{}& \frac{1}{2} V(H_*) (\partial^\nu h_{\mu\nu} - \frac{1}{2} \partial_\mu h )
+ \frac{\partial^2 V(H_*)}{\partial H^{\mu\nu} \partial H^{\rho\sigma}} 
\partial^\nu h^{\rho\sigma} = 0,
\nonumber\\
&{}& R_{\mu\nu} - \frac{1}{2} g_{\mu\nu} R 
= \frac{1}{2} V(H_*) ( \frac{1}{2} \eta_{\mu\nu} h - h_{\mu\nu} )
-  \frac{\partial^2 V(H_*)}{\partial H^{\mu\nu} \partial H^{\rho\sigma}} h^{\rho\sigma},
\label{Generalized Eq.2}
\end{eqnarray}
where we have used (\ref{Potential}) and for simplicity the Einstein's tensor 
$G_{\mu\nu} \equiv R_{\mu\nu} - \frac{1}{2} g_{\mu\nu} R$ is not expanded around
the Minkowski metric.

Then, a closer inspection reveals that there exist only two classes of potentials
which yield massive gravity without non-unitary modes. The strategy for finding
appropriate potentials is to require that with a suitable choice of the potentials
the linearized equations of motion (\ref{Generalized Eq.2}) reduce to a set of equations
\begin{eqnarray}
\partial^\nu h_{\mu\nu} - \partial_\mu h &=& 0,
\nonumber\\
R_{\mu\nu} - \frac{1}{2} g_{\mu\nu} R 
&=& \frac{m^2}{2} ( \eta_{\mu\nu} h - h_{\mu\nu} ),
\label{Massive Gravity}
\end{eqnarray}
which are the same as those of Fierz-Pauli massive gravity \cite{Fierz}.

Incidentally, the comparison between Eq. (\ref{Generalized Eq.2}) and Eq. (\ref{Massive Gravity})
clearly shows why the 't Hooft model for massive gravity encounters the problem associated with
the ghost mode. With the choice of $V = \Lambda$ in the 't Hooft model, Eq. (\ref{Generalized Eq.2})
become 
\begin{eqnarray}
\partial^\nu h_{\mu\nu} - \frac{1}{2} \partial_\mu h &=& 0,
\nonumber\\
R_{\mu\nu} - \frac{1}{2} g_{\mu\nu} R 
&=& \frac{\Lambda}{2} ( \frac{1}{2} \eta_{\mu\nu} h - h_{\mu\nu} ).
\label{'t Hooft model}
\end{eqnarray}
Since the coefficients in front of $h$ in the both equations are not 1 but $\frac{1}{2}$,
the decoupling of the ghost mode $h$ is not allowed in the 't Hooft model.
However, once the potentials with the higher derivative terms are included in the
action, we can adjust the value of the coefficients at will in order to get the desired value. 

One choice of such potentials is given by
\begin{eqnarray}
V(H_*) &=& 0,
\nonumber\\
\frac{\partial^2 V(H_*)}{\partial H^{\mu\nu} \partial H^{\rho\sigma}}
&=& - \frac{m^2}{2} ( \eta_{\mu\nu} \eta_{\rho\sigma} - \eta_{\mu(\rho} \eta_{\sigma)\nu} ),
\label{Potential1}
\end{eqnarray}
where we have defined $\eta_{\mu(\rho} \eta_{\sigma)\nu} \equiv \frac{1}{2}
( \eta_{\mu\rho} \eta_{\sigma\nu} + \eta_{\mu\sigma} \eta_{\rho\nu} )$.
It is easy to check that the Chamseddine and Mukhanov model is a special case of this
more general class of models when the potential $V$ is fourth order in $H^{AB}$.

Here let us comment on the sign of cosmological constant. To do that, note first
that the cosmological constant $\Lambda$ is equal to the value of the potential
$V$ at $H^{AB} = 0$. Thus, in the potential (\ref{CM Potential}), the cosmological 
constant reads
\begin{eqnarray}
\Lambda = - \frac{D(D-1)}{16} m^2,
\label{CC1}
\end{eqnarray}
which is negative for $D > 1$. A natural question then arises whether
the negative cosmological constant is an essential ingredient for producing
mass for gravitons as in Ref. \cite{Kaku2}. This question has been already
answered by taking account of a different potential in Ref. \cite{Chamseddine}.

The potential proposed for this purpose in  \cite{Chamseddine} is of form in
a general $D$-dimensional space-time
\begin{eqnarray}
V = - \frac{m^2}{4} [ \{ (\frac{1}{D} H)^2 - 1 \}^2 
\{ \alpha (\frac{1}{D} H)^2 - \beta \} - \tilde H_{AB} \tilde H^{AB} ],
\label{CM Potential3}
\end{eqnarray}
where $\alpha, \beta$ are some constants. It is straightforward to check that
this potential satisfies the equations (\ref{CM Potential2}) and (\ref{Potential1})
when we take $\alpha - \beta = \frac{D(D-1)}{4}$. In this model, the cosmological
constant is given by
\begin{eqnarray}
\Lambda = \frac{\beta}{4} m^2.
\label{CC2}
\end{eqnarray}
Hence, depending on the sign of $\beta$, the cosmological constant $\Lambda$
can become negative, zero or positive, which is an interesting aspect in this class
of potentials. 

The other interesting choice of potentials is given by
\begin{eqnarray}
V(H_*) &=& 2 m^2 \ne 0,
\nonumber\\
\frac{\partial^2 V(H_*)}{\partial H^{\mu\nu} \partial H^{\rho\sigma}}
&=& - \frac{m^2}{2} \eta_{\mu(\rho} \eta_{\sigma)\nu}.
\label{Potential2}
\end{eqnarray}
This class of massive gravity theories is new, so we shall present the simplest
example. Let us consider the case that the potential is a quadratic function
of $H^{AB}$, that is,
\begin{eqnarray}
V(H^{AB}) = \Lambda + c_1 \eta_{AB} H^{AB} + c_2 H_{AB} H^{AB},
\label{Example}
\end{eqnarray}
where $\Lambda$ is the cosmological constant, and $c_1, c_2$ are constants to be 
determined shortly. It is worthwhile to recall that with $\Lambda < 0$ and $c_2 = 0$,
this model reduces to that of 't Hooft, for which there is a non-unitary ghost
mode \cite{'t Hooft}. The condition (\ref{Potential}) gives rise to
\begin{eqnarray}
c_1 + 2 c_2 = \frac{1}{2} [ \Lambda + ( c_1 + c_2) D ].
\label{Example1}
\end{eqnarray}
Moreover, the equations (\ref{Potential2}) produce the relations
\begin{eqnarray}
\Lambda + ( c_1 + c_2 ) D &=& 2 m^2,
\nonumber\\
c_2 &=& - \frac{m^2}{4}.
\label{Example2}
\end{eqnarray}
As a result, the potential is of form
\begin{eqnarray}
V(H^{AB}) = \Lambda + \frac{3}{2} m^2 \eta_{AB} H^{AB} - \frac{1}{4} m^2 H_{AB} H^{AB},
\label{Example3}
\end{eqnarray}
where the cosmological constant takes the form $\Lambda = ( 2 - \frac{5}{4} D ) m^2$,
which is negative for $D > 1$. We conjecture that in this class of potentials,
the cosmological constant might be always negative since the 't Hooft model belongs
to this class.

\section{Conclusions and Discussion}

In this paper, we have generalized the Chamseddine and Mukhanov massive 
gravity model \cite{Chamseddine} to an arbitrary $D$-space-time dimension and showed the
unitarity of the model by gauge-fixing diffeomorphisms in two steps. 
It was found that the non-unitary mode, which is just the trace part of 
gravitational fields, does not appear in the physical propagating modes
by residual diffeomorphism.   

We have also contructed a general model with an arbitrary potential
which is a generic function of the internal induced metric $H^{AB}$. By requiring
the existence of the vacuum solution, which is a flat Minkowski space-time plus 
$\phi^A = \delta_\mu^A x^\mu$, as a classical solution, we find a constraint
on the potential. Furthermore, we can restrict the form of the potential
by requiring that the general model should coincide with the Fierz-Pauli
massive gravity \cite{Fierz}. 

The whole equations imposed on the potential turn out to have only two classes of nontrivial
solutions. One class of the solutions includes the Chamseddine and Mukhanov massive 
gravity model \cite{Chamseddine} as a specific choice of the potential and has an interesting
property that the cosmological constant takes any signs. The other class of 
the solutions is a new massive gravity model. As the simplest example, we have presented a
massive gravity model of this class with quadratic terms in $H^{AB}$ 
and shown that it has a negative cosmological constant.
We can conjecture that the appearance of the negative 
cosmological constant is an inherent feature of this class of potentials for causing
the gravitational Higgs mechanism. 

As a future work, we wish to construct a different sort of new massive gravity 
models and explore its physical implications. In this respect, it is worthwhile to notice that
one can construct a different induced metric $Y_{\mu\nu} = \eta_{AB} \nabla_\mu \phi^A 
\nabla_\nu \phi^B$ which was considered in Ref. \cite{Kaku2}. Thus it seems to be
natural to try to make new massive gravity models with an arbitrary function
of this metric $Y_{\mu\nu}$. We would also like to report this work in near
future.

\vs 1   


\begin{thebibliography}{99}

\bibitem{Percacci}
R. Percacci, {"The Higgs Phenomenon in Quantum Gravity",
\NP{\bf B353} (1991) 271, arXiv:0712.3545 [hep-th]};
C. Omero and R. Percacci, {"Generalized Nonlinear Sigma Models in Curved Space
and Spontaneous Compactification", \NP{\bf B165} (1980) 351.}

\bibitem{Kaku1}
Z. Kakushadze and P. Langfelder, {"Gravitational Higgs Mechanism", 
\MPL{\bf A15} (2000) 2265, arXiv:hep-th/0011245.}

\bibitem{Porrati}
M. Porrati, {"Higgs Phenomenon for 4-D Gravity in Anti de Sitter Space",
JHEP{\bf 0204} (2002) 058, arXiv:hep-th/0112166.}

\bibitem{Leclerc}
M. Leclerc, {"The Higgs Sector of Gravitational Gauge Theories", \AP{\bf 321} 
(2006) 708, arXiv:gr-qc/0502005.}

\bibitem{'t Hooft}
G. 't Hooft, {"Unitarity in the Brout-Englert-Higgs Mechanism for Gravity",
arXiv:0708.3184 [hep-th].}

\bibitem{Kaku2}
Z. Kakushadze, {"Gravitational Higgs Mechanism and Massive Gravity",
\IJMP{\bf A23} (2008) 1581, arXiv:0709.1673 [hep-th]; "Massive Gravity in Minkowski 
Space via Gravitational Higgs Mechanism", \PR{\bf D77} (2008) 024001, 
arXiv:0710.1061 [hep-th].}

\bibitem{Maeno}
M. Maeno and I. Oda, {"Classical Solutions of Ghost Condensation Models", 
\MPL{\bf B22} (2009) 3025, arXiv:0801.0827 [hep-th]; "Massive Gravity in Curved 
Cosmological Backgrounds", \IJMP{\bf A24} (2009) 81-100, arXiv:0808.1394 [hep-th].} 

\bibitem{Rubakov2}
V.A. Rubakov and P.G. Tinyakov, {"Infrared-modified Gravities and Massive
Gravitons", arXiv:0802.4379 [hep-th].}

\bibitem{Riess}
A.G. Riess et al., {"Observational Evidence from Supernovae for an Accelerating 
Universe and a Cosmological Constant", Astron. J. {\bf 116} (1998) 1009,
arXiv:astro-ph/9805201.}

\bibitem{Perlmutter}
S. Perlmutter et al., {"Measurements of Omega and Lambda from 42 High Redshift 
Supernovae", Astron. J. {\bf 517} (1999) 565,
arXiv:astro-ph/9812133.}

\bibitem{WMAP}
WMAP Collaboration, D.N. Spergel et al., {"Wilkinson Microwave Anisotropy Probe (WMAP) 
Three Year Results: Implications for Cosmology", Astron. J. Suppl. {\bf 170} 
(2007) 377, arXiv:astro-ph/0603449.}

\bibitem{Oda}
I. Oda, {"Gravitational Higgs Mechanism with a Topological Term", 
Adv. Studies Theor. Phys. {\bf 2} (2008) 261, arXiv:0709.2419 [hep-th].}

\bibitem{Oda1}
I. Oda, {"Strings from Black Hole", \IJMP{\bf D1} (1992) 355; K. Akama and I. Oda,
"Topological Pregauge Pregeometry", \PL{\bf 259} (1991) 431;
K. Akama and I. Oda, "BRST Quantization of Pregeometry and Topological Pregeometry",
\NP{\bf B397} (1993) 727.}

\bibitem{Chamseddine}
A.H. Chamseddine and V. Mukhanov, {"Higgs for Gravitons: Simple and Elegant Solution", 
arXiv:1002.3877 [hep-th].}

\bibitem{MTW}
C.W. Misner, K.S. Thorne and J.A. Wheeler, {"Gravitation", W H Freeman and 
Co (Sd), 1973.}

\bibitem{Fierz}
M. Fierz and W. Pauli, {"On Relativistic Wave Equations for Particles of Arbitrary 
Spin in an Electromagnetic Field", Proc. Roy. Soc. Lond. 
{\bf A173} (1939) 211.}

\end{thebibliography}
\end{document}